Original Paper

# Development and Design of FLKit: A Structured Onboarding Toolkit for Federated Learning in Health and Life Sciences

Ashkan Pirmani[*,1,2,3], Ilse Vermeulen[*,2,3], Goran Vinterhalter[1], Lotte Geys[2,3], Axel Faes[2,3], Muhammad Quamber Ali[1], Nishkala Sattanathan[4], Geert Vandeweyer[4], Yves Moreau[1], Liesbet M. Peeters[2,3]

[1]STADIUS, Department of Electrical Engineering (ESAT), KU Leuven, Leuven, Belgium

[2]Biomedical Research Institute (BIOMED), Hasselt University, Hasselt, Belgium

[3]Data Science Institute (DSI), Hasselt University, Hasselt, Belgium

[4]Department of Medical Genetics, University of Antwerp, Antwerp, Belgium

[*]Shared First Author

## Abstract

**Background:** Federated learning (FL) allows institutions to train models collaboratively without centralizing data, making it well suited for health and life sciences research operating under strict data protection regulations. Despite growing technical maturity, teams entering federated projects face a fragmented landscape of frameworks, governance requirements, and professional roles, with no structured starting point tailored to their specific context or background.

**Objective:** We present FLKit, an open, community-maintained onboarding toolkit designed to guide multidisciplinary teams through the full FL lifecycle in health and life sciences, addressing the absence of a role-aware, lifecycle-grounded entry point for practitioners from clinical, legal, governance, and technical backgrounds.

**Methods:** FLKit was developed using a design science approach modeled on the ELIXIR Research Data Management Kit (RDMkit). Development was organized across three tiers: a multidisciplinary core team covering governance, infrastructure, data wrangling, and model development; a wider consortium providing milestone-based review and roadmap input; and external practitioners interviewed to calibrate the toolkit against real-world practice.

**Results:** FLKit is organized around four lifecycle stages (Governance, Infrastructure, Wrangling, and Analysis), 11 role-specific entry points covering clinical researchers, data stewards, legal advisors, and research software engineers, a curated cross-disciplinary glossary, a reusable FAIR-aligned FL Story template for documenting and planning federated projects, and a directory of tools, frameworks, and communities. Since its demo launch in December 2024, FLKit has grown to 39 pages across eight content sections. Seven FL Stories document completed and ongoing projects across multiple sclerosis disability prediction, inflammatory bowel disease, genomics, and brain-computer interfaces.

**Conclusions:** FLKit addresses a structural gap in the FL landscape by providing a modular, role-aware entry point that helps multidisciplinary teams navigate governance, infrastructure, data harmonization, and analysis challenges without requiring prior expertise across all domains. It is openly available at https://uhasselt-biomedicaldatasciences.github.io/federated-learning-toolkit/ and actively welcomes contributions from across the life sciences.

## Introduction

The generation of health data has accelerated considerably over the past decade. Electronic health records, biobanks, disease registries, and genomic cohorts now span thousands of institutions across multiple jurisdictions. Yet the promise of large-scale, collaborative research that this data landscape implies remains largely unrealized. At the heart of health data science lies a persistent tension: the datasets most valuable for training robust, generalizable models are precisely those that are most sensitive, most heavily regulated, and hardest to move [1]. Data protection legislation, including the General Data Protection Regulation (GDPR) in Europe and the Health Insurance Portability and Accountability Act (HIPAA) in the United States, imposes strict constraints on the centralization and transfer of personal health data, making traditional pooled-data approaches practically out of reach for many research applications [2,3].

Federated learning (FL) offers a principled way to work within these constraints. First described by McMahan et al. [4], FL trains models locally at each participating institution and shares only model updates, rather than the underlying sensitive data or raw inputs. This approach preserves data sovereignty, supports regulatory compliance, and opens the door to collaborative research at scales previously unachievable. Reviews of FL applications in healthcare document growing evidence of feasibility across radiology, genomics, clinical prediction, and pharmacovigilance [1,5-8]. Tools including Flower [9], Fed-BioMed [10], and OpenMined (PySyft) [11] now provide mature orchestration infrastructure. Privacy-preserving techniques such as differential privacy [12,13] and secure aggregation [14] are increasingly integrated into FL pipelines for health data [15].

What the field has not addressed systematically is the entry problem. And that problem is worth spelling out, because it is not one challenge but several stacked on top of each other. Running an FL project requires coordinating components that span entirely different disciplines: distributed computing and networking to connect participating sites and protect communications; machine learning algorithms capable of learning effectively from data that is spread unevenly across nodes; mathematical privacy mechanisms such as differential privacy and secure multiparty computation; operating system-level containerization to deploy reproducible training environments across heterogeneous hardware; and data harmonization pipelines to bring locally held datasets into alignment with shared schemas before any model training can begin. On top of this technical stack sits a governance layer: legal agreements between institutions, ethics approvals, data access policies, and defined processes for model ownership and reuse. For every team that successfully deploys a federated pipeline, others stall at earlier stages, not because the technology failed, but because the people involved could not find where to begin.

The guidance landscape does not make this easier. Framework documentation for tools such as Flower or Fed-BioMed is written for experienced engineers already comfortable with distributed systems. Legal guidance such as the emerging European Health Data Space [16,17] addresses the regulatory landscape but offers little operational advice for the researcher or data steward planning their first federated study. Attempts to apply FAIR data principles [18] to federated settings have raised awareness of data governance requirements, but these principles do not yet translate into practical onboarding for the full range of roles involved. The result is a fragmented entry experience: technical documentation, legal frameworks, and community resources all exist, but they are scattered across disconnected communities, each written for a different subset of the audience and each assuming different prior knowledge.

The problem is also one of professional distance. FL demands simultaneous contributions from practitioners who operate in almost entirely different knowledge worlds. A federated model developer optimizing a distributed training algorithm and a data protection officer drafting a data protection impact assessment share the same project but almost none of the same vocabulary, tools, or conceptual reference points. Between them sit data stewards handling OMOP common data model mapping [19], infrastructure engineers configuring secure networks and container orchestration, clinical researchers interpreting model outputs, and project managers keeping institutions aligned across time zones. Each role has specific pain points, and guidance written for one does not readily transfer to another. This gap is documented [20]: a recent scoping review found that legal and governance dimensions remain among the least well-addressed aspects of FL in medical research, even as the technical literature has grown substantially, meaning that the users with the most critical cross-cutting responsibilities are getting the least support.

To address this gap, we developed the Federated Learning Toolkit (FLKit), an open, community-maintained resource and modeled on the ELIXIR Research Data Management Kit (RDMkit) [21]. RDMkit has served a similar onboarding function for research data management across the European life sciences community, and its success motivated a comparable approach for FL. FLKit is not a technical manual, a regulatory guide, or a training course. It is a structured entry point organized around the lifecycle of a federated project and the professional roles involved, designed to help users orient themselves, find the right resources for their context, and begin contributing responsibly.

In this paper, we describe the rationale, design process, and structure of FLKit, and reflect on the lessons learned in building a community-led toolkit for a fast-evolving and interdisciplinary domain. We discuss the challenges of serving diverse user roles, embedding ethical and FAIR-aligned practices from the outset, and sustaining a living resource beyond its initial development phase.

## Methods

FLKit was developed using a design science research framework, which treats the creation of practical artifacts (tools, methods, models) as a legitimate form of scholarly contribution when those artifacts demonstrably address a real-world problem. The core problem motivating FLKit was the entry problem described above: the absence of a structured, role-aware resource to guide users through the full FL lifecycle in health and life sciences contexts. This problem was identified through observation of common stalling patterns in federated projects, informal practitioner consultations during the design phase, and a review of available FL resources across framework documentation, community platforms, and regulatory guidance.

### Inspirational Framework

The design of FLKit was explicitly modeled on the ELIXIR RDMkit. RDMkit organizes its guidance around five complementary navigation dimensions: a staged data lifecycle (Plan, Collect, Process, Analyse, Preserve, Share, Reuse); role-specific entry points for data stewards, principal investigators, researchers, policy makers, and research software engineers; domain-specific guidance covering more than twenty research communities from cancer data to marine metagenomics; task-based pages addressing common RDM challenges from GDPR compliance to metadata documentation and licensing; and curated tool assemblies linking vetted resources to specific institutional workflows. This layered structure lets users enter through whichever dimension is most relevant to their current need and navigate outward from there, without having to absorb the entire resource before finding actionable guidance. As of 2025, RDMkit has grown to 138 pages, more than 250 contributors, and over 620 curated tools and resources, is recommended in the Horizon Europe Program Guide as the reference resource for data management good practices in the life sciences and has been formally documented as a replicable open research infrastructure.

FLKit applies the same structural logic to a more complex coordination problem. Whereas RDMkit guides a researcher or institution through managing their own data across its lifecycle, a federated project must coordinate work across institutional boundaries while each dataset remains at its source. FLKit therefore adds an explicit governance layer covering participation agreements between partners, shared decision-making on model architecture and training strategy, audit mechanisms for distributed model updates, and the handling of shared models after a project ends. These concerns are mapped onto a lifecycle of four stages (Governance, Infrastructure, Wrangling, and Analysis) that follows the practical sequence of decisions a federated team works through, together with a reusable FL Story template for documenting completed and planned projects.

### Structural Design Decisions

Three core design decisions shaped FLKit's architecture. First, we organized content around a four-stage FL lifecycle (Governance, Infrastructure, Wrangling, and Analysis). These stages were derived from a systematic analysis of the steps involved in a typical federated health data project, from initial governance planning through infrastructure deployment, data harmonization, and collaborative model training and evaluation. The lifecycle framing provides a stable conceptual scaffold that is deliberately decoupled from any specific tool or platform, so that when a framework updates its API or a regulatory document is superseded, the relevant FLKit page requires a link update rather than a reconceptualization of its underlying structure. Figure 1 shows the four stages and the representative concerns addressed at each.

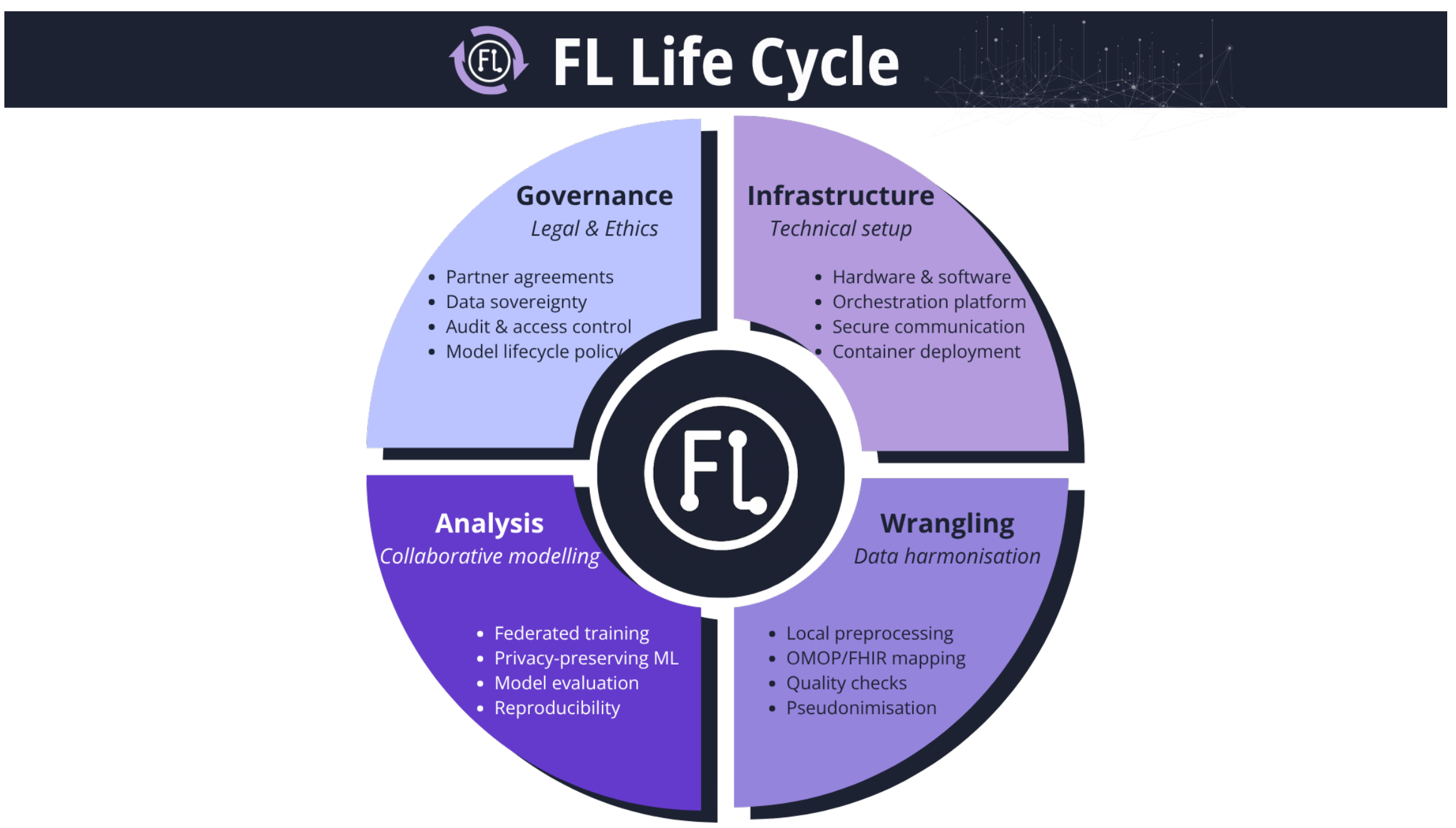

*Figure 1- The FLKit federated learning lifecycle, organized into four stages (Governance, Infrastructure, Wrangling, and Analysis), with representative concerns addressed at each stage.*

Second, we developed role-specific entry points covering the 11 professional roles most commonly involved in federated projects in health and life sciences contexts. This decision reflects the core insight that FL is inherently multidisciplinary and that the challenges of a data protection officer differ substantially from those of a federated model developer or a clinical researcher. Generic resources that address "the user" tend to serve no one particularly well. Each role entry point describes responsibilities, common challenges, recommended tools, and relevant lifecycle sections in terms appropriate for that role's background.

Third, we designed a FAIR-aligned FL Story template to document real-world use cases from federated projects. FL Stories serve two functions: they provide concrete examples that help newcomers understand what federated collaboration looks like in practice, and they contribute to a growing library of citable, structured accounts of federated projects in health and life sciences. The template was designed to capture project context, technical setup, governance arrangements, lessons learned, and pointers to tools and publications, in alignment with FAIR principles for research documentation.

### Development Process

Development was organized across three concentric layers of involvement. The innermost layer was a core development team whose composition was deliberately multidisciplinary, covering federated model development and analysis, software and systems engineering, project management and coordination, and governance. This spread ensured that design decisions in each lifecycle stage were shaped by the practitioner most directly accountable for that domain in a real federated project, rather than being specified by a single role and reviewed by others after the fact.

The second layer was a wider consortium team drawn from the broader FL and health data community. This group engaged at defined milestones and contributed to roadmap planning, providing cross-institutional perspective on whether the lifecycle structure, role taxonomy, and emerging content reflected the realities of federated projects beyond the core team's own context. Progress was reported at each milestone, and structural decisions were revisited when consortium input revealed gaps or misalignments.

The third layer consisted of external individuals who had no prior involvement in FLKit's development. These users were interviewed to provide an independent assessment of whether the toolkit was navigable, whether the role descriptions matched their experience, and whether the lifecycle stages reflected how federated projects actually unfold in their settings. Their input was used to calibrate the toolkit against practical ground truth and to surface blind spots that internal and consortium reviewers, by virtue of familiarity, were less likely to catch.

The toolkit itself was built as a website hosted on GitHub Pages, making it freely accessible and operationally straightforward to maintain. GitHub serves as both the hosting and contribution infrastructure, with two contribution pathways available: a pull-request-based route for those comfortable with version control, and a lower-barrier contact-based route for those who are not, so that domain experts in clinical, legal, or governance roles can contribute without needing software development experience.

## Results

FLKit was launched publicly as a demo in December 2024, with its current stable version released in November 2025, and has since grown to 39 pages organized across eight content sections. The toolkit is openly available at https://uhasselt-biomedicaldatasciences.github.io/federated-learning-toolkit/.

The eight sections correspond to distinct entry points into the toolkit as shown in Figure 2, each serving a different user need: the FL Life Cycle, the FL Glossary, the Frameworks directory, FL Stories, FL Communities, All Tools and Resources, All Training Resources, and Your Role. This organization means a user can enter through whichever dimension is immediately relevant to them, whether they need to understand the project lifecycle, identify their role-specific responsibilities, find a tool, or learn from a real-world example, without having to navigate through material designed for a different audience first.

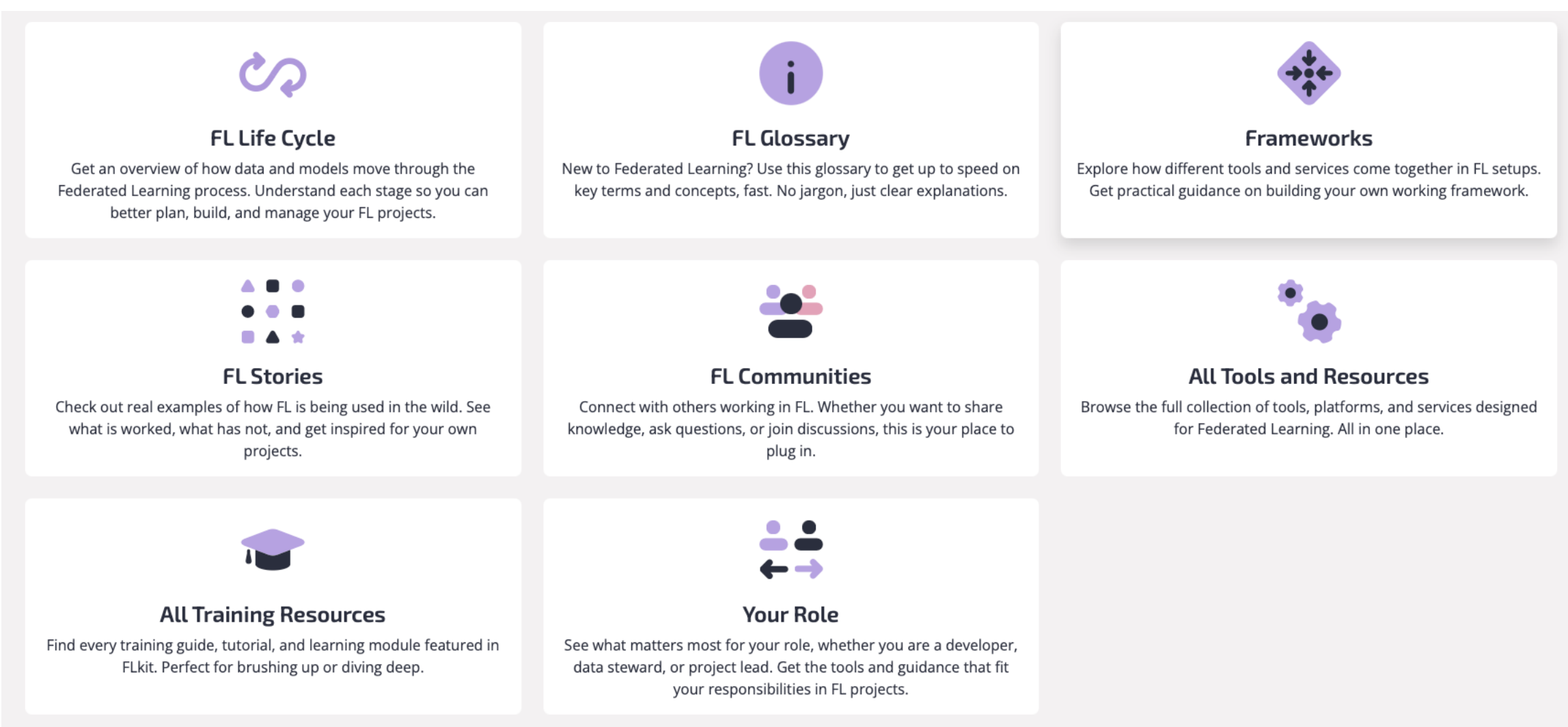

*Figure 2- Landing page of FLKit showing the eight content sections available to users.*

## The FL Life Cycle

The FL Life Cycle is the conceptual backbone of FLKit. It organizes the process of running a federated project into four sequential stages, each with its own page structured around three consistent questions: what this stage involves, why it matters, and what users should specifically consider. This consistent page architecture was a deliberate design choice, ensuring that a user entering any lifecycle stage encounters the same logical progression from definition to rationale to actionable guidance.

The Governance stage addresses the institutional, legal, and ethical arrangements that must be in place before any federated project can proceed. Each participating institution must understand who can do what with which data and models, under which conditions, and with which safeguards. The stage covers participation agreements that formalize partner responsibilities and liabilities, decision-making structures for model design and training strategy, legal compliance and institutional data policies, data sovereignty principles ensuring data remains under local control, audit trails and access control mechanisms, and model lifecycle governance covering how shared models are updated, shared, validated, and eventually retired. FAIR alignment is treated as a governance-level obligation rather than a downstream afterthought, with metadata, governance documents, and audit records expected to meet FAIR criteria.

The Infrastructure stage covers the technical systems that make local model training and cross-site coordination possible without moving data. This means hardware and software at each node capable of running training workloads, orchestration platforms to coordinate distributed rounds, secure communication between nodes using encryption in transit and at rest, containerized deployment environments using Docker and Kubernetes to ensure reproducibility across heterogeneous hardware, monitoring and backup mechanisms, and adherence to FAIR principles through standardized metadata and persistent identifiers. The stage addresses a frequently underestimated risk: that sites entering a federated project have very different infrastructure maturity levels, and that mismatched capacity can block the entire collaboration even when the governance and data preparation are in place.

The Wrangling stage addresses the preparation of locally held data for federated use. Because raw health data is messy, inconsistently coded, and structured differently across institutions, this stage is site-local but must achieve cross-site harmonization without central coordination. Guidance covers local preprocessing pipelines shared as version-controlled scripts or containers, mapping to common data models such as OMOP and FHIR for semantic alignment, data quality checks targeting missing values, unit inconsistencies, outliers, and duplicates, anonymization and pseudonymization procedures to comply with privacy requirements, and transformation transparency through documentation and versioning of all wrangling steps. The stage also addresses distributed validation approaches for checking harmonization across sites without pooling data, reflecting the practical challenge that quality problems at one node can corrupt the entire federated training round.

The Analysis stage is where FL delivers its primary value: collaborative model development and evaluation without data leaving its source. Guidance covers model architecture selection for federated settings, training strategies including federated averaging and the choice between synchronous and asynchronous rounds, handling of non-identically distributed data across sites which is a persistent technical challenge in real health data federations, integration of privacy-preserving techniques such as differential privacy and secure multiparty computation, evaluation using both local and federated metrics with attention to fairness and representativeness, model interpretability and communication to clinical and non-technical stakeholders, and documentation of analysis protocols and model configurations for reproducibility. Each consideration is linked to relevant tools and, where appropriate, to FL Stories that illustrate how the choice was handled in a real project.

### FL Stories and the FL Story Template

FL Stories form FLKit's primary mechanism for grounding lifecycle guidance in real-world practice. Seven completed or ongoing federated projects are documented as FL Stories at the time of writing, spanning a range of clinical and biomedical domains. “The Federated Learning Pipeline” story provides a hands-on implementation tutorial using the Flower framework and PyTorch, covering client and server configuration for practitioners building their first federated pipeline. “FL4E” documents the Federated Learning for Everyone framework [22], which was designed to lower the barrier to federated participation in clinical research by abstracting orchestration complexity. The GDSI story documents a three-layer federated data analysis pipeline developed for the Global Data Sharing Initiative in multiple sclerosis research [23]. The “FL-MS” story documents an application of FL to real-world multiple sclerosis data from the MSBase registry [24], exploring optimal FL configurations and personalization strategies across a large number of geographically distributed sites. The “FL IBD” story is a tutorial on using federated and deep learning models to predict complex phenotypes from genetic data in inflammatory bowel disease. The “WiNGS” story is a guide to querying federated genomic data using the WiNGS REST API [25], covering authentication, data access, and federated analysis for genomics researchers. The seventh, the “FBTTR-BCI” story, documents a federated block-term tensor regression approach for ECoG-based finger movement decoding [26].

The FL Story template is the structural scaffold that makes these stories comparable and citable. The current version of the template (v1.3) organizes any federated project into four stages with explicit structure, role assignments, and realistic timeline estimates. The Planning stage, which typically takes one to six months, covers problem definition and federation justification, ethics approval and legal compliance, stakeholder identification and team formation, and project timeline and resource planning. This stage is assigned primarily to principal investigators, legal teams, and ethics coordinators. The Preparation stage, typically two to twelve weeks, covers data exploration and quality assessment, infrastructure setup and security configuration, framework selection and testing, and data preprocessing and standardization, and is led by data teams and DevOps engineers. The Training stage, ranging from weeks to months depending on project complexity, covers algorithm design and model architecture, federated training implementation, integration of privacy-preserving techniques including differential privacy and secure multiparty computation, and performance evaluation and validation, and is led by machine learning engineers and security teams. The Deployment stage, typically two to eight weeks, covers reproducibility documentation and code sharing, model maturity assessment and validation, deployment planning and production readiness, and results dissemination and knowledge transfer, and involves research teams and institutional leadership. Each stage is explicitly labeled with its functional character, Foundation, Technical, Research, and Strategic respectively, making the template usable not only for documenting completed projects but as a planning scaffold for new ones.

### Role-Specific Entry Points

The "Your Role" section provides 11 entry points, each written for a specific professional identity rather than for a generic user. The documented roles are: Data Steward; Data Protection Officer and Legal Advisor; Clinical Researcher and Data Generator; Federated Model Developer and Data Scientist; Coordinator and Project Manager; Federated Data Steward; Policy Maker; Principal Investigator; Researcher; Research Software Engineer; and Trainer. Each role page follows a consistent structure covering the responsibilities relevant to that role within a federated project, common challenges specific to that professional context, and pointers to the most relevant lifecycle stages and tools. The role taxonomy spans the full spectrum from those who never touch model code (legal advisors, policy makers, governance coordinators) to those who operate almost entirely within it (federated model developers, research software engineers), reflecting the design principle that every practitioner on a federated team deserves guidance written in their own professional register, not a translation of someone else's technical manual.

### FL Glossary

Terminology in FL is neither settled nor uniform across disciplines. A term that is standard vocabulary for a machine learning engineer may be opaque to a clinical researcher, and governance concepts familiar to a legal advisor have no direct equivalent in the machine learning literature. FLKit's FL Glossary addresses this through a filterable table covering seven domain categories: FL, Bioinformatics, Data and Privacy, General machine learning, Clinical and Healthcare, Analytics, and Security and Privacy. The categories allow each user to scope the glossary to terms relevant to their background, rather than scrolling through definitions from adjacent fields. Entries range from FL-specific concepts such as robust aggregation, asynchronous FL, and client clustering to data governance terms covering anonymization, consent management, and common data models, to clinical concepts such as biomarker, case mix, and cohort study, to bioinformatics vocabulary including batch effect and allele frequency. The cross-disciplinary scope of the glossary is itself a design statement: it reflects the fact that federated projects require users to communicate across these domains, and that a shared vocabulary is a prerequisite for doing so effectively.

### FL Communities

FLKit's Communities section is a curated directory of active forums, events, and networks where FL users engage, share knowledge, and coordinate development. The section covers eight currently listed communities spanning different engagement formats and audiences. The Flower ecosystem is represented through three entry points: the Flower Slack and web forum community maintained by Flower Labs for users of the open-source Flower framework, the Flower Open-Source Federated AI Meetup organizing virtual and in-person events, and the Flower AI Summit, described as the world's largest FL conference and featuring research and industry talks alongside poster sessions. The OpenMined community is listed in two forms: the broader OpenMined open-source community focused on privacy-preserving machine learning, and the OpenMined Slack, which had over 17,000 members at the time of writing, making it one of the largest active communities in privacy-enhancing technologies. Additional entries include the IBM FL Slack for users working with IBM's FL framework, the NVIDIA Developer Forums covering FL topics related to NVIDIA FLARE, and a dedicated FL LinkedIn Group with approximately 950 followers providing a professional networking channel. The inclusion of a community directory reflects the recognition that no toolkit can remain current in a fast-moving field without practitioners who are embedded in those communities and can identify when FLKit's content needs to be revised.

### Frameworks, Tools, and Training Resources

Three further content sections complete FLKit's reference layer. The Frameworks section provides structured guidance on how tools and services combine into working FL setups, going beyond a listing of individual products to address framework selection and assembly decisions that depend on a project's specific governance, infrastructure, and analysis requirements. The All Tools and Resources section is a curated directory spanning orchestration platforms, common data model resources, containerization tools, and privacy-enhancing technologies, organized to reduce the search effort required when a team moves from lifecycle guidance to concrete implementation. The All Training Resources section aggregates tutorials, guides, and learning modules referenced across FLKit pages into a single access point, so that a practitioner who identifies a skill gap at any lifecycle stage can locate learning material without navigating away from the toolkit. Together these three sections function as the implementation bridge between FLKit's structural guidance and the broader ecosystem a practitioner will need to engage with as their project progresses

## Discussion

### Design Rationale: Curation over Comprehensiveness

A recurring observation from users entering FL projects is that the available resources, while numerous, are poorly connected. Framework documentation, legal guidance, governance templates, and training materials exist in separate communities and repositories, each assuming a level of prior orientation that newcomers rarely possess. FLKit was designed to address this not by creating new content from scratch, but by providing a structured layer of navigation and context above the existing ecosystem. This curatorial approach, inspired by RDMkit and its success within ELIXIR, reflects a deliberate philosophy: do not replicate what already exists, but help users find the right resource for their situation.

The lifecycle-based architecture provides a stable scaffold deliberately decoupled from any specific tool or platform. The FL tool landscape changes rapidly: frameworks are updated, new orchestration platforms emerge, and community standards shift. By anchoring guidance to lifecycle stages and roles rather than specific implementations, FLKit can accommodate this change without wholesale revision of its core structure. This separation is consequential for long-term maintenance: when a framework releases a breaking version change or a regulatory document is superseded, the relevant FLKit page requires a link update, not a reconceptualization of its underlying framework.

It is worth stating explicitly where FLKit departs from the model it draws on. RDMkit addresses how a single institution manages its own data responsibly, a coordination problem that is fundamentally internal. FL requires something categorically different: coordinating data use, model development, and decision-making across institutional boundaries without data ever leaving its source. This adds a governance dimension with no direct counterpart in general research data management, and it is this dimension that FLKit's lifecycle architecture is specifically designed to address.
The FL Story template is arguably FLKit's most structurally novel contribution. Where RDMkit links projects to published outputs and registered datasets, an FL Story captures what rarely surfaces in publications: the governance arrangements, orchestration choices, and privacy strategies that determined how a federated project actually unfolded. By providing a reusable FAIR-aligned template that makes these accounts comparable and citable, FLKit begins to build a library of federated project knowledge that is currently scattered across grey literature, internal reports, and tacit expertise.

The role-specific entry points represent what we consider a novel organizational contribution relative to existing FL resources. FL demands collaboration across clinical, legal, governance, and technical domains, and the challenges of a data protection officer navigating data protection impact assessment requirements differ substantially from those of a data scientist handling heterogeneity within data distributions. By providing role-appropriate language, common challenges, and targeted pointers for each entry point, the toolkit reduces cognitive overhead for newcomers and helps multidisciplinary teams identify where their responsibilities begin and end.

Developing the role taxonomy highlighted a persistent design tension: specificity versus breadth. Guidance specific enough to be actionable for one role risks being inaccessible or irrelevant to another, while guidance generic enough to serve all roles risks being too abstract to be useful for any of them. The solution FLKit adopts is to provide role-specific entry points into a shared body of lifecycle guidance: each role page describes responsibilities and common challenges in role-appropriate language, then links outward to the technical or governance lifecycle sections most relevant for that role. This allows depth to exist in the lifecycle sections while keeping the role pages accessible. The granularity of the role taxonomy itself required careful judgment. Too few roles would flatten meaningful distinctions: a data protection officer and a federated governance coordinator are both broadly non-technical but operate with entirely different concerns, vocabularies, and institutional responsibilities. Too many roles risk fragmenting guidance to the point where users cannot identify which applies to them. The 11 roles in FLKit represent a working hypothesis rather than a settled taxonomy, and community feedback will be essential for refining it over time.

A related gap that future versions should address is the experience of practitioners who occupy multiple roles simultaneously, which is common in smaller institutions or early-stage projects where a single person may serve as both data steward and project coordinator. Guidance that assumes clean role boundaries creates friction for these users. A quick-start pathway that helps users identify their primary responsibilities and navigate across multiple role pages accordingly may be a useful addition in future iterations.

### Embedding FAIR Principles and Ethical Practice

A design choice that distinguishes FLKit from many FL resources is the consistent embedding of FAIR principles and ethical considerations across all lifecycle stages and role pages, rather than treating these as a separate section to be consulted before or after the technical guidance. This reflects a view that responsible and FAIR-aligned FL is not achieved by appending a checklist at the end of a project, but by building the relevant considerations into each stage of the workflow from the start.

In practice, this means each lifecycle stage includes guidance on how FAIR principles apply in that specific context. The Governance stage covers persistent identifiers and citable governance documents. The Infrastructure stage addresses standardized metadata schemas and reproducible deployment environments. The Wrangling stage covers common data model alignment for semantic interoperability and documentation of preprocessing steps for reuse. The Analysis stage covers sharing analysis protocols, model configurations, and code in version-controlled, accessible formats. Similarly, ethical considerations including data sovereignty, consent management, data protection requirements, and risks of model bias are surfaced at the lifecycle points where they are most actionable rather than confined to a standalone ethics section that practitioners might bypass.

This integrated approach does introduce editorial risk: each section becomes more complex, and some users seeking purely technical guidance may find the ethical and legal caveats disorienting. Calibrating the depth of FAIR and ethics guidance for each audience and lifecycle stage remains an ongoing challenge. It may be worth exploring whether a fast-track pathway for technically experienced users could foreground implementation guidance while preserving FAIR and ethics material as clearly signposted supplementary content, rather than woven throughout.

### Community Sustainability

The most significant long-term challenge for FLKit is not technical but social: how to build and sustain a contributor community large and diverse enough to maintain quality, ensure breadth of coverage, and prevent the toolkit from reflecting only the perspectives of its founding team. This challenge is common to open community resources in research infrastructure. The experience of projects such as RDMkit, the FAIR Cookbook [27], and bio.tools suggests that community ownership requires deliberate investment in governance, contributor experience, and identity, not only content quality.

Lowering the barrier to contribution is necessary but not sufficient. Sustained engagement requires that contributors feel ownership over the resource and that their contributions are acknowledged, for example through ORCID-linked authorship in the contributor record, and that the editorial process be transparent and responsive. The tension between editorial quality control and openness to contribution is real: too little moderation risks inconsistency and loss of user trust; too much risks discouraging contributions and concentrating knowledge in a small group. FLKit's current model positions editors as moderators rather than content dictators, ensuring consistency and factual accuracy without imposing a single institutional perspective. As the community grows, formalizing this governance through a lightweight editorial board with rotating membership across institutions and roles will become important for both quality and legitimacy.

Several recruitment strategies appear promising based on analogous community resources. Embedding FLKit into training and onboarding programs, for example within community training events or FL workshops, creates a pipeline of users who may become contributors over time. FL Stories and use case submissions provide a low-stakes contribution pathway that does not require deep technical knowledge, making them particularly accessible for clinical researchers and project managers. Targeted outreach to existing FL communities, through framework user forums, working groups, and health data networks, can reach practitioners who are already invested in the field but may not yet be aware of FLKit.

### Content Currency in a Fast-Moving Field

FL is one of the fastest-evolving areas in applied machine learning and health data science. New frameworks emerge, established ones undergo significant revision, and regulatory guidance shifts as practitioners accumulate real-world experience. The most important structural defense against obsolescence is FLKit's curatorial model: the toolkit links to and contextualizes external resources rather than reproducing their content. When a tool's documentation is updated or a regulatory document superseded, the primary content changes at its source, and FLKit's role is to maintain an accurate pointer and a correct characterization of that resource's scope and relevance. This places a considerably lower maintenance burden on FLKit editors than a model in which the toolkit attempts to be the canonical source of truth for, say, how to configure Flower or what the GDPR says about federated data processing.

Nevertheless, even a curatorial model requires active maintenance. We recommend that FLKit adopt a structured review cycle, for example an annual community-wide content audit supplemented by quarterly link-checking supported by automated tooling. Community members who use specific tools or operate within specific regulatory contexts are often best placed to flag when content has become outdated, and encouraging this lightweight maintenance contribution could distribute the burden across a wider group. Making the toolkit's temporal scope explicit also matters: where guidance is tied to a specific tool version or regulatory instrument, this should be clearly signaled through version tags or last-reviewed dates. Transparency about currency is preferable to false confidence.

### Limitations

Several limitations should be acknowledged. The current content reflects the perspectives of a relatively small group of contributors, predominantly based in European research infrastructure contexts. Regulatory environments outside Europe, such as HIPAA-governed settings in the United States, certain data types with specific governance requirements such as genomic and imaging data, and low- and middle-income country settings with different infrastructure constraints are underrepresented. Expanding geographic and disciplinary breadth is both a sustainability goal and a content quality imperative.

FLKit has not yet been formally evaluated for usability or impact. While the development process incorporated stakeholder consultations and iterative feedback, a systematic assessment of how users navigate the toolkit, which sections they find most useful, and whether engagement translates into improved practices in federated projects has not been conducted. Such an evaluation, combining web analytics, user surveys, and qualitative interviews with teams that have used FLKit during a federated project, would provide valuable evidence for future development priorities and would strengthen the evidence base for community toolkits as an intervention in research infrastructure. We consider this a priority for the next development phase.

The question of long-term sustainability beyond initial project funding also deserves explicit attention. Many community resources in research infrastructure have struggled to maintain momentum after the conclusion of initiating projects, and FLKit is not immune to this risk. Embedding FLKit within an established infrastructure such as ELIXIR and distributing governance responsibility across multiple institutions and communities are important mitigations. Exploring models in which maintenance is supported through contributions-in-kind from benefiting institutions, for example through secondments of data stewards or developers, may offer a more durable approach than dependence on project-based grants.

### Conclusions

FLKit addresses a genuine and growing need. As FL moves from research prototype to practical approach for multi-institutional health data collaboration, users who must implement it require structured, role-aware, and ethically grounded guidance spanning the full lifecycle of a federated project. By providing a curated, modular, and community-maintained entry point into this complex landscape, FLKit aims to lower the barrier to responsible federated collaboration and help teams navigate the most common challenges of governance, infrastructure, and data harmonization.
The experience of developing FLKit suggests that the answer to sustaining shared infrastructure lies not in trying to be comprehensive or self-contained, but in being strategically curatorial: curating the right links, the right roles, the right lifecycle structure, and the right contribution pathways, and investing in the governance and community processes that keep those curation decisions honest over time. We invite researchers, data professionals, clinicians, legal experts, and infrastructure specialists across the FL community to contribute to FLKit. The toolkit will be as good as the community that sustains it.

## Acknowledgements

We would like to thank all the experts who generously contributed their insights and expertise during the interviews. This work was supported by the Research Foundation - Flanders (FWO) for ELIXIR Belgium (I000323N).

## Conflicts of Interest

None declared.

## Abbreviations

FL: Federated Learning
FLKit **:** Federated Learning Toolkit
RDMkit **:** Research Data Management Kit